\documentclass[acmsmall]{acmart}

\AtBeginDocument{%
  \providecommand\BibTeX{{%
    \normalfont B\kern-0.5em{\scshape i\kern-0.25em b}\kern-0.8em\TeX}}}

\setcopyright{acmcopyright}
\copyrightyear{2020}
\acmYear{2020}
\acmDOI{0000001.0000001}

\usepackage{multirow}
\usepackage{color}
\usepackage{subfigure}
\usepackage{makecell}
\usepackage{booktabs}
\usepackage{caption}

\newcommand{\zhengyang}[1]{\textcolor[rgb]{0.3,0.7,0.7}{#1}}
\newcommand{\sidong}[1]{\textcolor[rgb]{0.1, 0.5, 0.32}{{#1}}}

\begin{document}

\title{From Lost to Found: Discover Missing UI Design Semantics through Recovering Missing Tags}


\author{Chunyang Chen}
 \affiliation{%
   \department{Faculty of Information Technology}
 	\institution{Monash University}
 	\city{Melbourne}
 	\country{Australia}}
 \email{chunyang.chen@monash.edu}

 \author{Sidong Feng}
 \affiliation{%
 	\department{Research School of Computer Science}
 	\institution{Australian National University}
 	\city{Canberra}
 	\country{Australia}
 }
 \email{u6063820@anu.edu.au}

 \author{Zhengyang Liu}
 \affiliation{%
 	\department{Research School of Computer Science}
 	\institution{Australian National University}
 	\city{Canberra}
 	\country{Australia}
 }
 \email{zhengyang.liu@anu.edu.au}

 \author{Zhenchang Xing}
 \affiliation{%
 	\department{Research School of Computer Science}
 	\institution{Australian National University}
 	\city{Canberra}
 	\country{Australia}
 }
 \email{zhenchang.xing@anu.edu.au}

 \author{Shengdong Zhao}
 \affiliation{%
 	\institution{National University of Singapore}
 	\city{Singapore}
 	\country{Singapore}
 }
 \email{zhaosd@comp.nus.edu.sg}

\renewcommand{\shortauthors}{Chen and Feng, et al.}

\begin{abstract}
    Design sharing sites provide UI designers with a platform to share their works and also an opportunity to get inspiration from others' designs.
    To facilitate management and search of millions of UI design images, many design sharing sites adopt collaborative tagging systems by distributing the work of categorization to the community.
    However, designers often do not know how to properly tag one design image with compact textual description, resulting in unclear, incomplete, and inconsistent tags for uploaded examples which impede retrieval, according to our empirical study and interview with four professional designers. 
     Based on a deep neural network, we introduce a novel approach for encoding both the visual and textual information to recover the missing tags for
     existing UI examples so that they can be more easily found by text queries. We achieve 82.72\% accuracy in the tag prediction. Through a simulation test of 5 queries, our system on average returns hundreds more results than the default Dribbble search, leading to better relatedness, diversity and satisfaction.	
    \end{abstract}

\setcopyright{acmcopyright}
\acmJournal{PACMHCI}
\acmYear{2020} \acmVolume{4} \acmNumber{CSCW2} \acmArticle{123} \acmMonth{10} \acmPrice{15.00}\acmDOI{10.1145/3415194}

\ccsdesc[500]{Human-centered computing~User interface design}
\ccsdesc[300]{Human-centered computing~User interface management systems}
\ccsdesc[100]{Human-centered computing~Graphical user interfaces}
\keywords{Tags, Graphical User Interface, Designers, Semantics, Deep learning}

\maketitle

\section{Introduction}

Graphical User Interface (GUI) is ubiquitous which provides a visual bridge between a software application and end users through which they can interact with each other.
Good GUI designs are essential to the success of a software application and can gain loyalty from the software users~\cite{jansen1998graphical}.
However, designing good GUI is challenging and time-consuming, even for professional designers. 
On one hand, GUI designers must follow many design principles and constraints, such as fluent interactivity, universal usability, clear readability, aesthetic appearance, consistent styles~\cite{galitz2007essential, clifton2015android, web:appleDesign}.
On the other hand, they also strive to be creative and come up with unique, original ideas.
One way to design good UI is to learn from each other, and as a result, many design sharing websites such as Dribbble~\cite{web:dribble}, Behance~\cite{web:behance}, Coroflot~\cite{web:coroflot}, etc. have emerged and gained significant popularity among designers' community.

Successful design sharing sites have become repositories of knowledge with millions of professional design artworks.
For example, Dribbble has over 3.6 million designs~\cite{web:dribbble2018}.
As most GUI designs are in image format which is difficult to organize or search, textual tags are one way that users sift through this volume of information.
When tags are integrated into platform tools, they can be used to filter and retrieve relevant content.
On Dribbble, tags can be used to set up the notification, personalize the design feed appearing on the homepage, annotate the design trend, and search for specific content.
Tags are also used internally for the algorithms for recommending a list of related or potentially interesting GUI designs.

Collaborative tagging systems~\cite{choi2018will, golder2006usage, macgregor2006collaborative} allow sites to distribute the work of categorization across the entire community. 
It has now been utilized by many sites to help organize large online repositories.
For example, on Dribbble, designers are asked to apply up to 20 tags to describe their uploaded GUI designs. 
Collaborative tagging systems empower users to actively participate in the organization of the community content and distribute the effort of categorizing content across many users.
As a result, more content can be labeled, making it easier to discover.

While tagging one UI design may seem relatively easy, prior work suggests that tagging involves a complex set of cognitive tasks~\cite{fu2008microstructures}.
Tagging with others in a collaborative tagging system involves sharing and negotiating concepts or understanding from many different people.
Collaborative tagging can be regarded as a collective sensemaking~\cite{golder2006usage} or a distributed cognitive system~\cite{fu2008microstructures}.
But some issues emerge during this collaborative tagging~\cite{golder2006usage}, especially in design sharing sites.
First, users may use different words to describe the same UI design based on their own background knowledge.
As there are no rules for how a tag applies to one UI design, the tags provided by different designers for even similar meaning can be different, resulting in a very organic vocabulary, especially for designers who are new to the website.
The increase in the use of incoherent tags may hinder content searching or navigation.
For example, when designers want to search for UI designs with ``user interface'', the UI design tagged with ``ui'' or ``uidesign'' will be omitted.

\begin{figure*}
	\centering
	\includegraphics[width=0.95\textwidth]{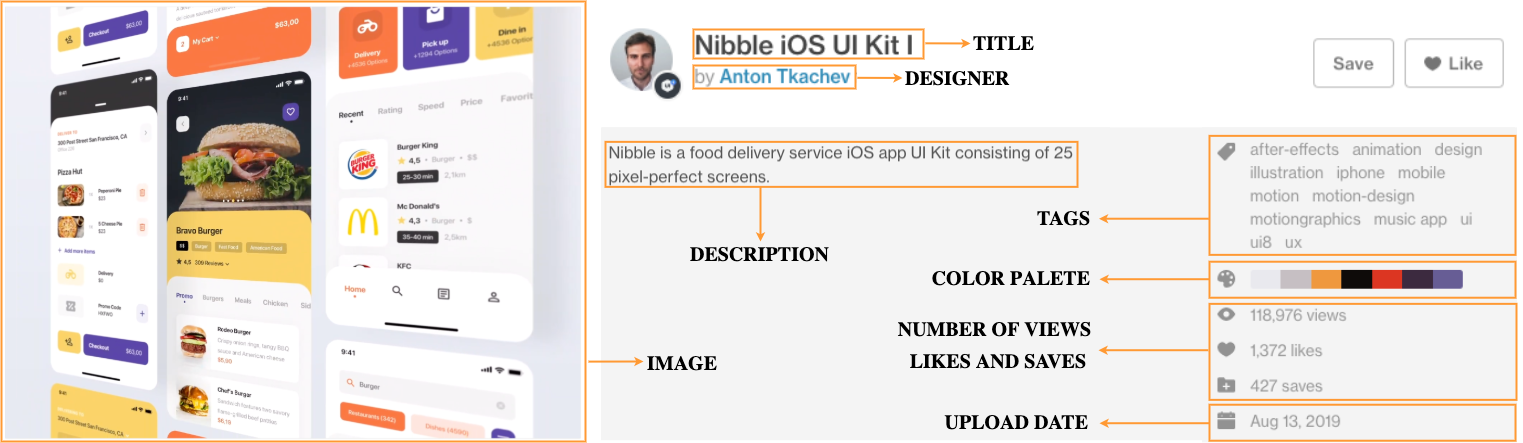}
	\caption{An example design from the design sharing website {\tt www.dribbble.com} of which tags illustrate the two problems with the tagging-based search.} 
	\label{fig:missingTag}
	\vspace{-10pt}
\end{figure*}

Second, users may extract different topics from the same design according to their own understanding, hence missing some closely related tags associated with the GUI design images. 
When tagging the UI design, users tend to provide only a few keywords that describe the most obvious visual or semantic contents, while content in other aspects may be simply omitted.
Such an issue is further aggravated due to the gap for transforming the visual GUI design information to textual tags.
For example, the design in Figure~\ref{fig:missingTag} shows the UI design of a food app. 
Although it contains a food view and a checkout page, this design isn't tagged with such related keywords, resulting in the potential failure of the retrieval.
Once designers search it with ``user interface'', it may not appear in the returning result, though ``ui'' is the abbreviation of it.

Therefore, in addition to users' tagging, we also need a more proactive mechanism  of tag assurance which could check the tag completeness of one UI design before it is uploaded, and remind the UI designers to fix issues if any. 
The goal of our work is to develop such a proactive policy assurance mechanism which can complement the existing collaborative tagging mechanism to augment existing tags of the GUI design, hence benefiting the GUI search and management in the design sharing sites. 
Our work is data driven and built on top of existing collaborative editing patterns by users. 
Therefore, our research questions are two-fold: 1) \textit{what kinds of issues are there with tagging in design sharing sites?} 2) \textit{does a deep-learning model help designers recover missing tags for UI designs?}

To answer the first research question, we conduct an empirical study of tags attached to tens of thousands of GUI designs on Dribbble.
By analyzing the existing tags, we find that users adopt incoherent tags with a large inconsistent vocabulary according to their own preferences.
But some frequently used tags which convey the main semantic of the GUI design images emerge including tags related to the GUI platform, color, software functionality, screen functionality, screen layout, etc.
Interviews with professional designers (UI design creators and uses of design sharing sites) from big companies also confirm the tag usage issues (i.e., inconsistency, tag missing).
Considering the wide range of missing tags, it would require significant manual effort to develop a complete set of rules for guiding collaborative tagging.

This challenge motivates us to develop a deep-learning based tag completeness assurance janitor to recommend to design owners or other users the potential missing tags that may be needed to augment existing tags. 
This janitor can also justify its prediction by pointing out the specific part of GUI designs relevant to the predicted tags. 
To recommend missing tags for UIs, we employ a hybrid approach that leverages both the visual and textual information.
We train a binary classifier with UI design images from Dribbble labeled with this tag as positive data, while UI design images labeled with other tags in the same category as negative data.
The trained model can help recommend tags for not only the new UI designs before sharing, but also remind designers about adding more tags for existing UI designs.
The experiments demonstrate that our model with the average accuracy as 0.8272 significantly outperforms the other machine learning based models.

To demonstrate the usefulness of our model, we also carry out a user study to show if our model for recovering missing tags can help retrieve more related UI designs compared with the default Dribbble search.
After analyzing ten participants' feedback, we find that our model can provide more related and larger number of UI candidates given the queries than the baselines.



The contributions of this work are threefold.
	\begin{itemize}
		\item We systematically investigate the collaborative tagging problems of the design sharing site including incoherent tag usage and missing tags. The informal interviews with professional designers also confirm these issues qualitatively.
		\item Based on the emerging tag categories, we develop a customized deep-learning based method for specifically recommending missing semantic tags to the existing design by leveraging both visual and textual information according to the UI design characteristics. We have released the source code, experiment results, and tag categorization to the public\footnote{\url{https://github.com/UITagPrediction/CSCW2020}} for replication and further extension.
		\item We conduct large-scale experiments to evaluate the performance and limitations of our approach. Our approach achieves good accuracy compared with baselines on a large-scale UI design dataset. The evaluation of our method for improving searching performance also demonstrates the usefulness of our model.
\end{itemize}

\section{Related Works}
Our work is problem driven, rather than approach driven.
As our study is to recommend missing tags for UI design images, we are introducing related works in three aspects, i.e., works for helping UI design, collaborative tagging and tag prediction.


\subsection{UI Design}
GUI provides a visual bridge between apps and users through which they can interact with each other. 
To assist the mobile UI design, many studies are working on large-scale design mining including investigating the UI design patterns~\cite{alharbi2015collect}, color evolution~\cite{jahanian2017colors, jahanian2017mining}, UI design testing including usability~\cite{zhao2020seenomaly, liu2020owl} and accessibility~{chen2020unblind}, UI decomposition~\cite{chen2020object} and design implementation~\cite{chen2018ui}.
Liu et al~\cite{liu2018learning} follow the design rules from Material Design to annotate the mobile GUI design to represent its semantics.
Swearngin et al~\cite{swearngin2018rewire} adopt the image processing method to help designs with converting the mobile UI screenshots into editable files in Photoshop, so that designers can take it as a starting point for further customization. 
To render inspirations to the designer, Chen et al~\cite{chen2019storyboard} propose a program-analysis method to efficiently generate the storyboard with UI screenshots, given one app executable file. 
Fischer et al~\cite{fischerimaginenet} transfer the style from fine art to GUI.
All of these works are helping simplify the design process for professional design.
In contrast, our study is to assist with searching the existing UI design to render designers some inspirations.

Retrieval-based methods~\cite{chen2019gallery, chen2020wireframe} are also used to develop user interfaces. 
Reiss~\cite{reiss2018seeking} parses developers' sketch into structured queries to search related UIs of Java-based desktop software in the database.
GUIfetch~\cite{behrang2018guifetch} customizes Reiss's method~\cite{reiss2018seeking} into the Android app UI search by considering the transitions between UIs.
Deka et al~\cite{deka2017rico} leverage auto-encoder to support UI searching by inputting the rough sketch.
Similar to Reiss's work~\cite{reiss2018seeking}, Zheng et al~\cite{zheng2019faceoff} parse the DOM tree of one user-created rough website to locate similar well-designed websites by measuring tree distance.
Note that all of these works either take the wireframe/sketch or partial source code as the input for searching the UI design.
Different from them, we are annotating UI design with high-level natural-language semantic tags which can help textual queries for UI design retrieval.

\subsection{Collaborative Tagging System}
Tagging is to summarize the content with several compact keywords~\cite{golder2006usage, macgregor2006collaborative} and collaborative tagging allows individual users to create and apply tags to online items.
Collaborative tagging systems are widely adopted by different sharing sites including Q\&A (e.g., Quora~\cite{web:quora}), picture sharing (e.g., Flickr~\cite{web:flickr}), web bookmarking sharing (e.g., Delicious~\cite{web:delicious}), etc.
Instead of central authority, collaborative tagging systems constitute an entirely new method of information organization, which allows the entire community to generate tags and label content, typically to make the content easier for themselves or others to find later.

Creating and applying tags in order to organize content requires the cognitive process of categorization~\cite{fu2008microstructures, fu2010semantic}.
Although tags provide a way for sites to organize and for users to search the content, there are still some problems with it.
So many research works have aimed to study on human tagging behavior.
For example, Suchanek et al~\cite{suchanek2008social} examine whether tags are applied accurately by humans. 
Multiple empirical studies~\cite{golder2006usage, halpin2007complex, wetzker2010tag} conduct to evaluate human categorization consistency.
Fu et al~\cite{fu2010semantic} study on whether users are influenced by others' tags.
The conflict and divisions in collaborative tagging have been further spotted~\cite{choi2018will}.
The empirical study in our work also confirms those tagging issues in design sharing sites.
But different from existing works which mostly focus on the empirical study of such phenomenon, we further develop a method to help recover the missing tags and keep the consistency of tags across the site.
Categorization of tags can analogously facilitate navigation and searching in the now-booming collaborative tagging system. 
A number of papers deal with exploring plenty of tag type categories for online resources of different kinds, such as book~\cite{golder2006usage}, movie~\cite{sen2006tagging}, TV program~\cite{melenhorst2007usefulness}, music~\cite{bischoff2008can}, blog~\cite{li2017types}.
To the best of our knowledge, our work is the first to build a UI tag categorization using data mining and semi-automatic techniques.

\subsection{Tagging Prediction}

Automatic image annotation (also known as automatic image tagging or linguistic indexing) is the process by which a computer system automatically assigns metadata in the form of captioning or keywords to a digital image~\cite{web:imageAnnotation}.
Most work has focused on personalized tag recommendation, suggesting tags to the user, bookmarking a new resource using collaborative filtering, taking into account similarities between users and tags~\cite{mishne2006autotag, xu2006towards, tso2008tag, liang2008collaborative}.
As pointed out in~\cite{eom2011improving} that the visual content of images are a valuable information source, some researchers propose conventional machine learning methods for tag recommendation based on manually-crafted features~\cite{sigurbjornsson2008flickr, chen2013fast, kalayeh2014nmf}. 

Apart from the image, the text attached to the image (e.g., existing tags) may be another resource for tag recommendation.
Murthy et al.~\cite{murthy2015automatic} explored a k-nearest-neighbor-based canonical correlation analysis (CCA) method for tag recommendation, which not only utilizes the convolutional neural network visual features but also explicitly incorporates the word embedding semantic information.
Ma et al.~\cite{ma2017image} proposed a joint visual-semantic propagation model to the tag image and introduced a visual-guided LSTM to capture the co-occurrence relation of the tags.
Bylinskii et al.~\cite{bylinskii2017understanding} recommended tags for infographics by adopting patch learning to select representative textual and visual elements from an infographic. Rawat et al.~\cite{rawat2016contagnet} proposed a context-aware model to integrate context information with image content for multi-label hashtag prediction.

But note that different from prior works about tag recommendations for natural images on social-media platforms (e.g., Flickr), we are specifically targeting at uncovering semantic tags of UI design images for assisting designers in effectively seeking others' UI design. 
Inspired by the existing works on tag recommendation, we customize the off-the-shelf model for modeling both visual and textual information in three aspects. 
First, we adopt binary classifiers to boost the performance of the classification model. 
Previous works formulating this tag-recommendation problem as a multi-class multi-label classification task, but it always leads to data imbalance issues~\cite{liu2019synthetic} as the occurrence number of each tag differs significantly.
Instead, according to our observation, most related (frequent) tags attached to the UI design are exclusive with each other, such as ``black vs red'', ``login page vs checkout page'', ``music app vs sport app''. 
Therefore, we develop a novel binary classifier for each tag and that method naturally solves the problem of data sparsity and imbalance.

Second, to support the construction of binary classifiers, we carry out an empirical study of existing tags of 61,700 UI designs for distilling exclusive tags. Based on a series of methods including association rule mining and merging abbreviations and synonyms, we obtain a categorization of most frequent UI-related tags. Tags within the same category are exclusive from each other, enabling the consecutive construction of binary classifiers.

Third, we also develop a data visualization method to help designers understand why we recommend certain tags for their UI designs. For each recommended tag, we draw a saliency map highlighting the conclusive features captured by our model. Compared with just a simple prediction, this explainability can help post owners or post editors judge the validity of the recommended post edits and may encourage them to accept the tool's recommendation. Making the prediction criteria more explicit is crucial for the acceptance of our deep learning based approach for tag recommendation.

\section{Empirical Study of Design Sharing}
\label{sec:empiricalStudy}
In this work, we select Dribbble as our study subject, not only because of its popularity and large user base, but also due to its support for collaborative tagging.
Dribbble is an online design sharing community which contains over 3.6 million uploaded artworks covering graphic design, user interface design, illustration, photography, and other creative areas.
Only professional designers can be invited to upload their design works for sharing with the community, leading to high-quality design within the site.
Since its foundation in 2009, Dribbble has become one of the largest networking platforms for designers to share their own designs and get inspirations from others'.
For each design work, the graphical design is attached with the corresponding metadata including a title, a short description, tags specified by the designer, and comments from other users as seen in Figure~\ref{fig:missingTag}.

To collect designs and associated metadata from Dribbble, we build a web crawler based on the Breadth-First search strategy~\cite{najork2001breadth} i.e., collecting a queue of URLs from a seed list, and putting the queue as the seed in the second stage while iterating.
Apart from the graphical design (always in png, jpeg format), our crawler also obtains the corresponding metadata including title, description, tags, uploading date, number of likes, saves, views, etc.
The crawling process continued from December 27, 2018 to March 19, 2019 with a collection of 226,000 graphical designs.
Note that these graphical designs are diverse including UI design, logo, icon, clip art, branding, typography, etc.
In this work, as we are concerned with the UI related design, we only select the UI attached with related tags such as ``user interface'', ``ui'', ``ux'', ``app design'', ``website design'', etc.
Finally, we obtain a dataset of 61,700 UI design and their metadata.
Based on this large dataset, we carry out an empirical study of collaborative tagging in Dribbble to understand its characteristics for motivating the required tool support.

\subsection{What are designers sharing?}

\begin{figure*}
	\centering
	\includegraphics[width=0.95\textwidth]{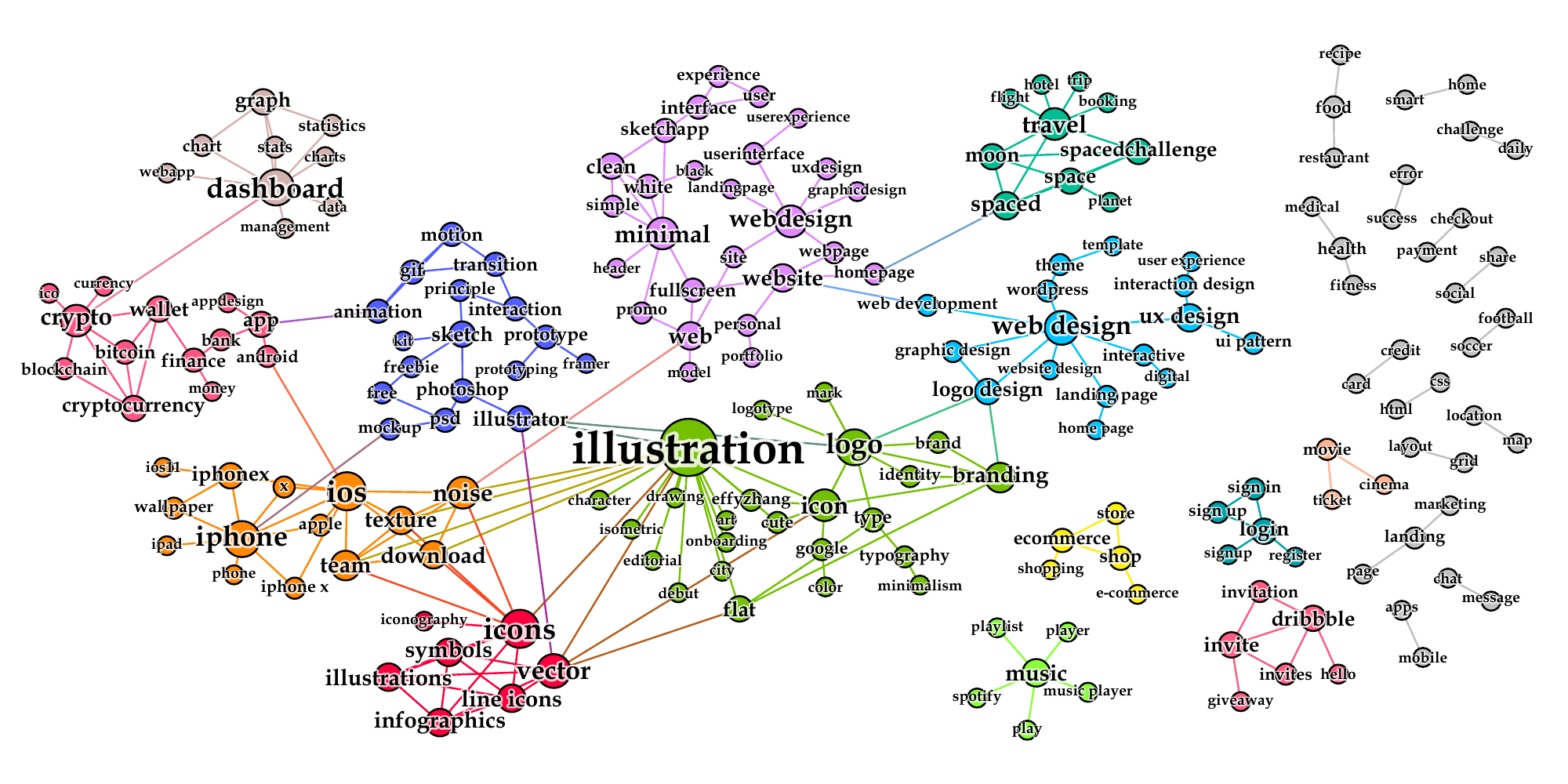}
	\caption{The UI-related tag associative graph from December 27, 2018 to March 19, 2019.}
	\label{fig:communitydetection}
	\vspace{-1.0em}
\end{figure*}

Within the Dribbble site, the design creators can add at most 20 tags for their design work.
These natural-language tags convey the semantics of the UI design such as internal structural information (e.g., ``dashboard'', ``list'', ``chart'') and belonging-software information (e.g., ``ecommerce'', ``food'', ``travel'').
To provide readers with a better understanding of detailed semantic embedded in the UI design, we adopt the association rule mining~\cite{agrawal1994fast} and community detection~\cite{blondel2008fast} for visualizing the landscape of collected tags.

\subsubsection{Overview of UI semantics}
We first collect all the UI designs with tags $t$, and tags of each design figure are correlated. 
In the example of Fig.~\ref{fig:missingTag}, \textit{iphone} is one kind of \textit{mobile} phones.
Each design is considered as a record and the design tags as the items in the record.
We use association rule mining~\cite{agrawal1994fast} to discover tag correlations from tag co-occurrences in designs.
As we want to mine the pair-wise correlation of tags and construct the landscape graph, we find frequent pairs of tags.
A pair of tags is frequent if the percentage of how many designs are tagged with this pair of tags compared with all the designs is above the minimum support threshold $t_{sup}$.
Given a frequent pair of tags $\{t_1, t_2\}$, association rule mining generates an association rule $t_1 \Rightarrow t_2$ if the confidence of the rule is above the minimum confidence threshold $t_{conf}$.
The confidence of the rule $t_1 \Rightarrow t_2$ is computed as the percentage of how many designs are tagged with the pair of tags compared with the designs that are tagged with the antecedent tag $t_1$.
Note that the original tags for filtering out non-UI designs like ``ui'', ``user interface'' are removed from the figure, as they are related to all tags in our dataset.

Given the mined tag association rules, we construct an undirected graph $G(V, E)$, where the node set $V$ contains the tags (i.e., technologies) appearing in the association rules, and the edge set $E$ contains undirected edges $<t_1, t_2>$ (i.e., tag associations) if the two tags have the association $t_1 \Rightarrow t_2$ or $t_2 \Rightarrow t_1$\footnote{The edge is undirected because association rules indicate only the correlations between antecedent and consequent.} i.e., the confidence of either $\{t_1, t_2\}$ or $\{t_2, t_1\}$ larger than the threshold.
Each edge has a confidence attribute indicating the strength of the technology association.
Considering the balance between the information coverage and overload, we set the $t_{sup}$ as 0.001 and $t_{conf}$ as 0.2, resulting in the 197 nodes and 246 edges.

We then carry out community detection~\cite{blondel2008fast} to group highly related nodes in the graph.
In graph theory, a set of highly correlated nodes is referred to as a community (cluster) in the network.
In this work, we use the Louvain method~\cite{blondel2008fast} implemented in the Gephi~\cite{bastian2009gephi} tool to detect communities.
We visualize different communities with different colors and the nodes with higher frequency as large size as seen in Fig.~\ref{fig:communitydetection}.

\subsubsection{Vocabulary of UI semantics}

\begin{table*}
\centering
\makebox[0pt][c]{\parbox{1\textwidth}{%
    \begin{minipage}[b]{0.49\hsize}\centering
        \scriptsize
        \caption{The categorization of some most frequent UI-related tags.}
        \label{tab:tagCategory}
        \begin{tabular}{>{}p{2.5cm} >{}p{3.8cm}}
        \toprule
        \bf{CATEGORY} & \bf{ASSOCIATED TAG NAME}\\
        \midrule
        \underline{PLATFORM} \\
    	\bf{\quad Website:} &  Website, Web, Mac, Macbook\\
    	\textbf{\quad Mobile:}  & Mobile, Phone, IOS, Iphone, Android\\
    	\textbf{\quad Tablet:} & Tablet, Ipad, Ipadpro\\
    	\underline{COLOR} \\
    	\textbf{\quad White:} & White \\
    	\textbf{\quad Yellow:} &  Yellow, Golden, Orange \\
    	\textbf{\quad Red:} &  Red \\
    	\textbf{\quad Pink:} &  Pink \\
    	\textbf{\quad Purple:} & Purple\\
    	\textbf{\quad Blue:} &  Blue, DarkBlue, SkyBlue \\
    	\textbf{\quad Green:} &  Green, DarkGreen, Aquamarine \\
    	\textbf{\quad Grey:} &  Grey, Silver, DarkGray\\
    	\textbf{\quad Brown:} &  Brown \\
    	\textbf{\quad Black:} &  Black\\
    	\underline{APP FUNCTIONALITY} \\
    	\textbf{\quad Music:} & Music, Musicplayer, MusicApp \\
    	\textbf{\quad Food \& Drink:} &  Food, Restaurant, Drink \\
    	\textbf{\quad Game:} &  Game, Videogame  \\
    	\textbf{\quad Health \& Fitness:} &  Fitness, Health \\
    	\textbf{\quad News:} &  News\\
    	\textbf{\quad Sport:} &  Sport, Gym, Workout\\
    	\textbf{\quad E-commerce:} &  E-commerce, Store, OnlineShop\\
    	\textbf{\quad Social Networking:} &  SocialNetwork, Blog, Messenger, Facebook, Instagram, Dating, Chat\\
    	\textbf{\quad Travel:} &  Travel, Trip, Tourism\\
    	\textbf{\quad Weather:} &  WeatherApp, Temperature\\
    	\textbf{\quad Lifestyle:} &  Fashion, Furniture, Real Estate\\
    	\textbf{\quad Education:} & Education, E-learning\\
    	\textbf{\quad Reference:} & Dictionary, Atlas, Encyclopedia\\
    	\textbf{\quad Entertainment:} &  Movie, TV, Netflix, Youtube\\
    	\textbf{\quad Medical:} &  Medical, Healthcare, Hospital \\
    	\textbf{\quad Books:} & DigitalReading, DigitalBookstroe\\
    	\textbf{\quad Kids:} &  Kids, Children\\
    	\textbf{\quad Finance:} &  Finance, Wallet, Bank, Business, Insurance, Marketing\\
    	\textbf{\quad Utilities:} &  Calculator, Clock, Measurement, WebBrowser\\
    	\textbf{\quad Navigation:} &  DrivingAssistance, TopographicalMaps, PublicTransitMaps\\
    	\underline{SCREEN FUNCTIONALITY} \\
    	\textbf{\quad Landing Page:} &  LandingPage\\
    	\textbf{\quad Login:} &  Login, Signin\\
    	\textbf{\quad Signup:} &  Signup, Registration\\
    	\textbf{\quad Checkout:} &  Checkout, Payment\\
    	\textbf{\quad Search:} &  Search\\
    	\textbf{\quad Profile:} &  Profile\\
    	\textbf{\quad Contact Page:} &  Contact, ContactPage\\
    	\underline{SCREEN LAYOUT} \\
    	\textbf{\quad Dashboard:} &  Dashboard\\
    	\textbf{\quad Form:} &  Form\\
    	\textbf{\quad Table:} & Table\\
    	\textbf{\quad List:} &  List \\
    	\textbf{\quad Grid:} & Grid\\
    	\textbf{\quad Gallery:} &  Gallery\\
    	\textbf{\quad Toolbar:} &  Toolbar, Toolbox\\
    	\textbf{\quad Chart:} &  Chart\\
    	\textbf{\quad Map:} &  Map, MapView\\
      \bottomrule
      \end{tabular}
    \end{minipage}
    \hfill
    \begin{minipage}[b]{0.49\hsize}\centering
      \scriptsize
      \caption{The 40 most frequent UI related tags with their abbreviations and synonyms and in brackets indicate the number of occurrences.}
      \label{tab:morphologicalForm}
      \begin{tabular}{>{\bfseries}p{2.3cm} >{}p{3.7cm}}
        \toprule
        \textbf{STANDARD(\#)} & \bf{ABBREVIATION \& SYNONYMS}\\
        \midrule
        ui (61309): & user interface, user\_interface, user-interface design, uidesign\\
    	website (28009): & web, websites, webpage, website development, web design\\
    	ux (24209): & user experience, uxdesign, ux\_design\\
    	mobile (8554): & mobiledesign, mobile phone, mobile\_design, smartphone\\
    	illustration (7159): & illustation, digital\_illustration, kids\_illustration \\
    	app (5887): & apps, application, app development, app design\\
    	landing page (5536): & landing-page, landingpage, landing page design\\
    	minimal (4938): &  minimalism, minimalistic, minimalist\\
    	ios (4741): & ios8, ios9, ios11, ios\_design \\
    	iphone (4736): & iphone x, iphonex, iphone 7, iphone\_8\\
    	icon (4230): & icons, icon design, icon pack\\
    	logo (3704): & logo design, logos, logotype\\
    	food (2881): & fastfood, food\_blog, junk\_food, doughnut\\
    	clean (2723):& clear, clean\_design\\
    	flat (2481): & flat design, flat-design, flat-art\\
    	interaction (2402): & interactive, microinteraction, interaction design, user interaction\\
    	dashboard (2141): & dashboard design, dashboards\\
    	branding (2071): & branding design, rebranding, selfbranding\\
    	sketch (2060): & sketching, sketches, adobe\_sketch \\
    	ecommerce (1974): & e-commerce, online commerce, shopping\\
    	vector (1940): & vectors, vector art\\
    	product (1841): & products, product page, product detail\\
    	typography (1820): & interface typography, 3d\_typography\\
    	gradient (1671): & gradients, gradient design, blue gradient\\
    	gif (1441): & gifs, looping\_gif\\
    	layout (1400): & layout design, layouts\\
    	concept (1378): & conceptual, concepts, concept art\\
    	motion (1361): & motion graphics, motion design\\
    	responsive (1347): & responsive design, response\\
    	music (1251): & music player, musician, musical, concert\\
    	restaurant (1221): & restaurants\\
    	profile (1204): & profiles, user\_profile, userprofile\\
    	travel (1197): & travelling, travel agency, travel guide\\
    	animation (1194): & animations, 3danimation, 2danimation\\
    	simple (1108): & simply, simplistic, simple\_design\\
    	graphic (1047): & graphics, graphic design, graphicdesigner\\
    	color (1000): & colors, colorful\\
    	white (988): & whitelabel, white design, white theme\\
    	login (919): & log\_in, sign\_in, login screen\\
    	modern (915): & modernistic, fashionable\\
        \bottomrule
      \end{tabular}
    \end{minipage}
}}
\end{table*}

As seen in Figure~\ref{fig:communitydetection}, many related tags are linked together such as (``travel'', ``hotel''), (``minimal'', ``simple''), (``ecommerce'', ``shop''), etc. 
To identify the set of frequently occurred UI tag categories, we adopted a consensus-driven, iterative approach to combine the observed tag landscape generated by our method with existing expert knowledge documented in books and websites such as \textit{Mobile Design Pattern Gallery}~\cite{neil2014mobile} and \textit{Google's Material Design}~\cite{web:googlematerial}.

We also performed an iterative open coding of 1,000 most frequent co-occurring tags with tag ``ui'' in Dribbble, or approximately 8.2\% of the dataset (12,244 in total).
Two researchers from our team independently coded the categories of these tags, noting any part of the initial vocabulary.
Note that both researchers have UI design experience in both mobile apps and websites.
After the initial coding, the researchers met and discussed discrepancies and the set of new tag categories until consensus was reached.

This process yielded 5 main semantic UI categories: PLATFORM, COLOR, APP FUNCTIONALITY, SCREEN FUNCTIONALITY, SCREEN LAYOUT.
Each main category also contains some sub-categories as seen in Table~\ref{tab:tagCategory}.
For example, the APP FUNCTIONALITY category contains ``MUSIC'', ``FOOD \& DRINK'', ``GAME'', and the subcategory ``FOOD \& DRINK'' contains UI design tagged with ``Food'', ``Restaurant'', ``Drink'', etc.

\subsection{Do designers use consistent tagging vocabulary?}
\label{sec:tagNormalization}
During the process of open coding the categories of UI semantic, we find that one tag can be written in different styles.
For example, the tag ``visualization'' not only has its standard format, but also its derivations including its synonyms like ``visualisation'', ``visualizations'', and even misspelling ``vizualization''.
As the UI design tagging process is informal and they are contributed by thousands of designers with very diverse technical and linguistic backgrounds, the same concept may be mentioned in many \textit{morphological forms} including \textit{abbreviations, synonyms, and misspellings}.
The wide presence of morphological forms poses a serious challenge to information retrieval and also vocabulary construction.
For example, the query ``ecommerce app'' may not find the UI design tagged with ``e-commerce'', ``e commerce design''.

To extract these morphological forms of the same words, we adopt a semi-automatic method~\cite{chen2017unsupervised} which leverages both the semantic and lexical information within the word.
We first train a word embedding~\cite{mikolov2013distributed} model to convert each tag into a vector that encodes its semantic.
The tag with semantic similarity will be spotted such as ``minimal'' and ``minimalistic'', ``ui'' and ``user interface'', ``travel'' and ``hotel''.
Then we define a lexical similarity threshold based on the string edit distance~\cite{levenshtein1966binary} to further check if two semantically-related words are similar enough in the form.
So the synonyms are extracted like ``minimal'' and ``minimalistic''.
To discriminate the abbreviations from semantically-related words, we define a set of rules e.g., the character of the abbreviation must be in the same order as they appear in the full name (ui, \textbf{u}ser \textbf{i}nterface), and the length of the abbreviation must be shorter than that of the full name, etc.

Although the automatic approach can generate many possible pairs of morphological forms, some of them may still be wrong.
Therefore, two researchers from our team manually check all of these pairs independently, and only take the morphological forms into consideration when the consensus is reached. 
Examples of some most frequent morphological forms are listed in Table~\ref{tab:morphologicalForm}. 

\begin{figure*}
	\centering
	\subfigure[Complete]{
		\includegraphics[width = 0.23\textwidth]{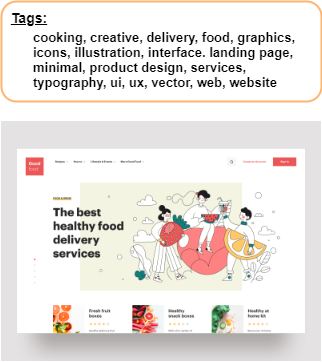}
		\label{fig:complete_1}}
	\hfill
	\subfigure[Complete]{
		\includegraphics[width = 0.23\textwidth]{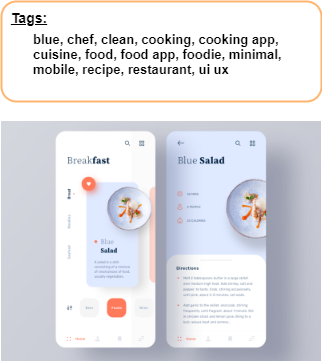}
		\label{fig:complete_2}}	
	\hfill
	\subfigure[Incomplete]{
		\includegraphics[width = 0.23\textwidth]{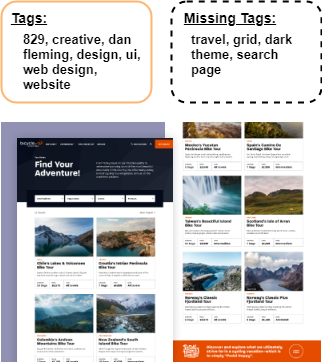}
		\label{fig:incomplete_1}}	
	\hfill
	\subfigure[Incomplete]{
		\includegraphics[width = 0.23\textwidth]{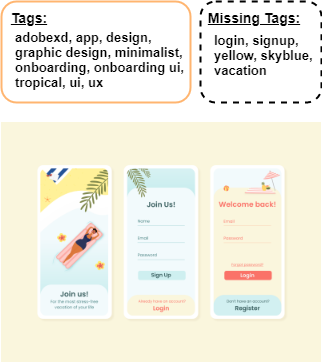}
		\label{fig:incomplete_2}}
	\vspace{-1.1em}
	\caption{Examples of UI examples with complete $\&$ incomplete tags.}
	\label{fig:UIexamples}
\end{figure*}
 
\subsection{How do designers share and search UI designs in the design sharing site?}
After quantitatively spotting the issues of collaborative tagging for the design sharing site, we further qualitatively get some feedback from informal interviews with four professional UI designers, including one interaction designer and one visual designer from Google, and the other two visual designers from Huawei.
These designers acknowledge that they regularly (at least once weekly) visit design sharing websites like Dribbble.
They appreciate the creativity and aesthetics of high-quality UI design artworks on Dribbble, which inspire their own design work. 
However, they also pinpoint several information needs in their work which demand us to think beyond existing content sharing mechanisms.

First, as the contributors of the Dribbble site, some designers do not pay much attention to their tagging, though it is crucial for the GUI index and searching.
One designer from Huawei mentioned to us that he did not intentionally use the semantic words (e.g., functionality, layout) to annotate his design works, as he did not know much what tags are there and which tags to use.
Instead, he attached his UI design with some fancy tags like ``2019trending'' to make his design more attractive to other designers.
A structured vocabulary of available tags is needed to assist designers in efficiently annotating their design works for other designers' retrieval.

Second, there is a gap between existing UI tags and the query due to two reasons.
On the one hand, some UIs are missing relevant semantic tags.
For example, one Google designer was designing a website about Yoga.
When she searched ``yoga``, there are many different kinds of returning results from Dribbble including the icons, charts, illustration which are not related to her designed UI or website.
She had to manually filter out irrelevant ones, and that process took much time.
But when she zoomed in a more specific query like ``yoga website``, only very few results were returned by Dribbble.
On the other hand, designers may adopt different words to tag the same GUI, resulting in the difficulty of retrieval.
Another designer in Google mentioned that he would use different words in the same meaning for searching such as both ``travel`` and ``trip`` for related GUI, but that process is inefficient as many results from two queries overlap, and he may miss some potential synonyms.
Some effective mechanisms have to be devised to help retrieve more results by not only adding more semantic tags to the GUI design, but also organizing existing tags.
Some example UIs with complete or incomplete tags from these designers can be seen in Figure~\ref{fig:UIexamples}.

\textbf{Summary:}
Design sharing site with diverse UI designs is crucial for inspiring professional designers.
The collaborative tagging provides a way for designers to search for their UI design of interest.
However, there are several limitations with current UI design tagging including using different same-meaning tags for the same UI design based on their own background knowledge, missing highly related tags for describing the UI design.
Such limitations may hinder their potential utility as navigational cues for searchers.
Therefore, an automated tag augmentation model is needed to recommend missing tags and keep existing tags consistent across different UI designs. 

\section{Augment Tags for the UI Design}
\label{sec:tagAugmentation}
Although extracting morphological forms mentioned in the last section can boost the performance of UI retrieval by normalizing the existing tags from designers, some UIs are still missing related tags which makes them unsearchable.
Therefore, a method to recover the missing tags of existing UIs is necessary.
However, based on the labeled data, we could not find a code-based heuristic that distinguishes tags from UI design images with high accuracy: the existing UI designs are too diverse.
So, we propose a hybrid deep learning method modelling both the visual and textual information to predict missing tags of the UI design. 

\subsection{Approach overview}
\begin{figure*}
	\centering
	\includegraphics[width=0.95\textwidth]{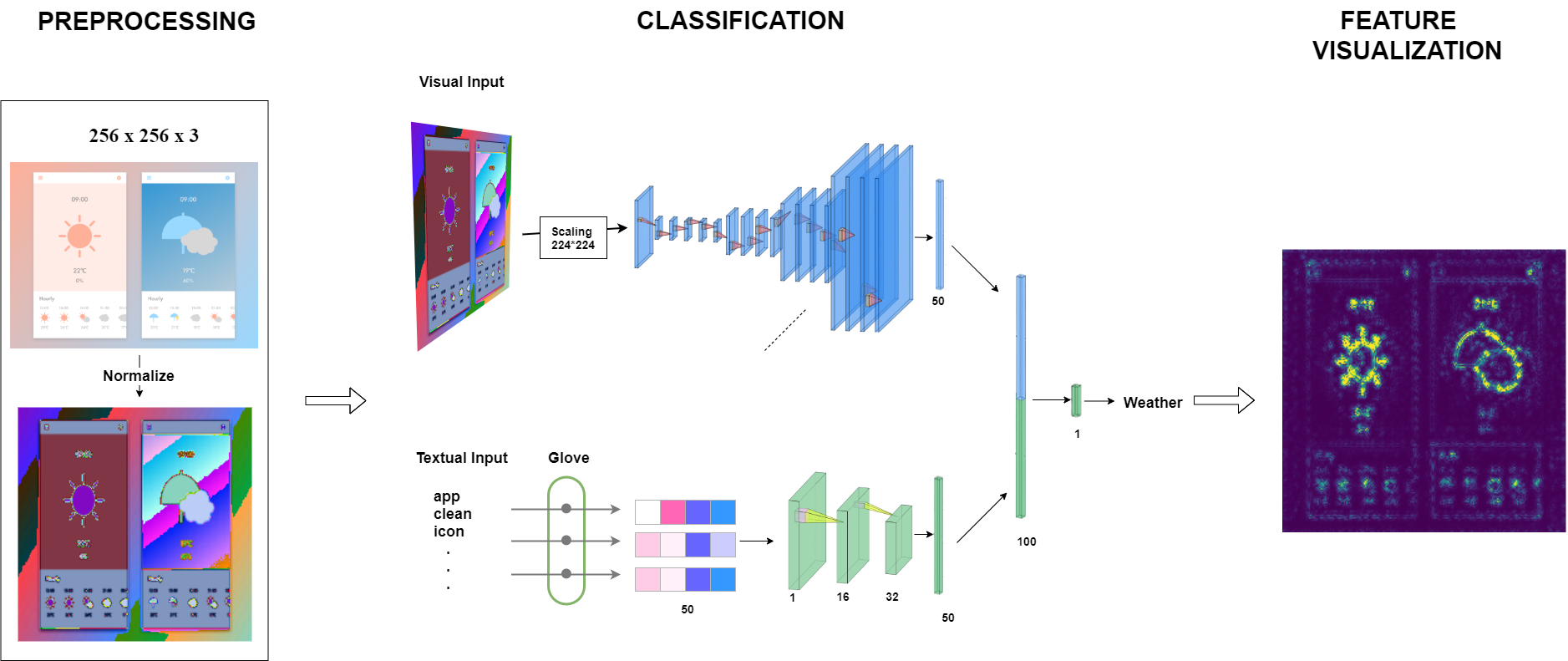}
	\caption{The architecture of our tag prediction model.}
	\label{fig:CNN_Architecture}
	\vspace{-5mm}
\end{figure*}

The overview of our approach can be seen in Fig~\ref{fig:CNN_Architecture}.
To train the deep learning for tag augmentation, we first collect all existing UI designs with specific tags identified in our empirical study (Section~\ref{sec:trainingData}). 
We then develop a tag augmentation model by combining a CNN model for capturing visual UI information and a fully-connected neural network for capturing textual information of existing tags (Section~\ref{sec:model}).
Additionally, to understand how our ensemble model makes its decisions through the visual and textual information, we apply a visualization technique (Saliency Maps~\cite{simonyan2013deep} and Softmax function~\cite{chen2018data}) for understanding which part of the figure and which words lead to the final prediction (Section~\ref{sec:visualization}).

\subsection{Dataset preparing}
\label{sec:trainingData}

We formulate this tag augmentation task as a classification problem.
We leverage the tag categorization during the creation of the UI semantic vocabulary as the label, and corresponding UI design images and attached tags as the training data.
Note that each UI design may own many semantic tags such as ``mobile'', ``blue'', ``social'', ``login'' and ``form'', and these tags are not exclusive from each other.
Therefore, we cannot regard all tags as equal for training a multi-class classifier for tag prediction.
In this work, we train a binary classifier for each tag label.
Such binary classifiers also benefit the system extensibility for new tags, as we only need to train a new binary classifier for the new tag without altering existing models for the existing tags.
For training each binary classifier, we take the UI design with that tag as the positive data.
We randomly take the same amount of UI designs attached with tags which are in the same category of the target tags in Table~\ref{tab:tagCategory} as the negative training data, as tags in the same categories are always exclusive from each other.
To keep the balance of data, we make the amount of positive data equal to that of negative data.
Therefore, the chance level for accuracy is 50\%.

We further preprocess the data in two aspects. 
For each tag, we merge similar tags based on the morphological/synonym form identification and some minor manual check.
For example, the tag ``traveling'', ``travel guide'', ``trip'' are all merged to tag ``travel''.
Note that the manual check is only carried out once for grouping similar tags in dataset preparation and no more human efforts are needed in the model construction.
For each image, we normalize the pixel value by dividing the mean and standard deviation so that each pixel value is in the range between 0 to 1 for accelerating the training process.

\subsection{Model Architecture}
\label{sec:model}
In this section, we propose an end-to-end deep fusion model to jointly learn visual and textual information. 
The architecture of the proposed network is shown in Fig~\ref{fig:CNN_Architecture}, which contains two main components.
One CNN model takes the GUI design image as the input for extracting the visual features.
The other CNN model takes the existing tags as the input for extracting the textual information.
We then merge both information sources to the final fully-connected layers for predicting the tag of the GUI design.


\subsubsection{Capturing Visual Information}
Convolutional Neural Network (CNN)~\cite{lecun1998gradient, krizhevsky2012imagenet} is a class of deep, feed-forward neural networks, most commonly applied to analyze visual imagery. 
The architecture of a typical CNN is structured as a series of stages such as different layers called convolution layers, pooling layers, and fully-connected layers.
In the convolutional layers, the convolution of an image is to use a kernel (e.g., $3\times3$ matrix) and then slide that window around the image.
At each position in the image that the kernel ($K$) visits, we multiply its values element-wise with the original matrix, then sum them up i.e., matrix dot multiply.
After the convolutional layers, it is the pooling layer.
Max Pooling works very much like convoluting, where we take a kernel and move the kernel over the image with no overlap.
It reduces parameter numbers and computation in the network, and hence controls the overfitting.

Different from the original CNN model, we adopt a more advanced one, ResNet (Residual Neural Network)~\cite{Resnet} with skip connections among the convolutional layers so that the model can be designed as very deep.
But due to the limitation of our dataset and deep structure of the model, we cannot train the ResNet directly with our small-scale dataset.
Therefore, we adopt the existing pretrained model which already distills the visual features from the ImageNet~\cite{imagenet}, the largest image dataset including more than 14 million natural images.
Based on the pre-trained ResNet, we further use our dataset to fine-tune the ResNet to adjust the parameters in different layers.

\subsubsection{Capturing Textual Information}
Apart from the GUI design picture, designers also attach some tags to index their picture.
According to the observation in Figure~\ref{fig:communitydetection}, we find that some tags co-occur frequently, indicating some implicit relations among them.
To predict the missing tags of a certain GUI design picture, we also take the existing tags into consideration in our model.
Given a set of existing tags, we first adopted the pre-trained word embedding model~\cite{web:glove} to convert each tag into a 50-dimension vector that encodes its semantic.
As an image has $n_{tags}$ number of tags, we concatenate the vectors of tags together, and obtain a $n_{tags}\times50$ vector representation.
Finally, we pad and synthesize all tags' vector representation of each image into a $50\times50$ map (similar format to a figure) and input it to a 2-layer CNN for extracting the salient textual information.
Note that we do not adopt the widely-used Recurrent Neural Network (RNN) to model tag information, as there is no explicit order among tags.


\subsubsection{Combining Visual and Textual Information for Prediction}
Based on the picture and tag input to the ResNet and text CNN, we obtain the embedding vectors which encode the visual and textual information.
We then concatenate both vectors and feed them into another fully-connected layer for the final prediction.
Therefore, we link both the ResNet, text CNN and final fully-connected layer as an end-to-end system for taking the UI design pictures and attached tags as the input, and output the predicted missing tags for them.

Note that we train a separate binary classifier for each tag.
We regard the UI is of one tag if the output is smaller than 0.5, while not if the output is equal to or larger than 0.5.
For detailed implementation, we adopt the max-pooling and use our dataset to fine-tune the last classification layer of the pretrained ResNet when modeling the visual information.
For modeling the textual data, we adopt the Glove word embedding~\cite{web:glove} with embedding size as 50.
To make our training more stable, We use a constant learning schedule with rate 0.0001 and adopt Adam as an optimizer~\cite{kingma2014adam}, BCELoss as the loss function.

\subsection{Prediction Visualization}
\label{sec:visualization}
One of the main drawbacks of the deep learning system is its interpretability i.e., how the model gives a certain prediction, especially considering the multi-channel data that we adopt as the input.
To gain insight into our classifier for the prediction results, we visualize the contribution from both the visual and textual input to the final prediction.

For visualizing the contribution from the visual input to the prediction, we adopt a visualization technique~\cite{simonyan2013deep} which calculates a saliency map of the specific class, highlighting the conclusive features captured by ResNet. 
Given the prediction result, the activities of decisive features in intermediate layers are mapped back to the input pixel space by class saliency extraction~\cite{simonyan2013deep}.
For one input UI design image, the final encoding of ResNet is mostly based on the weight of each pixel. 
Given the prediction result, the derivative weight map can be computed by back-propagation. Then the elements of weight are rearranged to form the saliency map.
As shown in the final part of Figure~\ref{fig:CNN_Architecture}, it can be seen that the ResNet model predicts the tag ``Weather'' for that input design due to the existence of the sun, umbrella, and cloud in the UI design image.

For visualizing the contribution from the textual input to the prediction, we first trace back through the CNN model to locating the filtered phrases in the input layer.
Then we predict the contribution score of the phrases' corresponding features in the fully connected layer to the prediction class, through a series of feature mapping, unpooling and deconvolution operation.
The predicted class of our CNN model is calculated by applying Softmax function~\cite{chen2018data} on the sum of the dot product of the weight matrix in the fully connected layer and the output feature vector of the concatenate layer.
For each UI design, we rank the score of each tag and select three top tags that contribute most to the prediction of the missing tag.
For example, it can be seen in Figure~\ref{fig:deepvisualization} that for the predicted tag "travel", the model highlights the words related to the landscape like "forest", "nature" and "mountain".

\section{Accuracy Evaluation}
\label{sec:AccurayEvaluation}

\subsection{Dataset}
Given all the data assigned to each tag, we prepare the dataset as discussed in Section~\ref{sec:trainingData}.
The foundation of the deep learning model is the big data, so we only select tags (including its derived tags) with frequency larger than 300 in Table~\ref{tab:tagCategory} for training the model.
Therefore, there are 26 tags (e.g., ``music'', ``signup''\footnote{The full list of tags and their frequency can be seen at \url{https://github.com/UITagPrediction/CSCW2020}}) left with the number of UI designs ranging from 312 to 13,233.
Note when splitting the dataset, we only carry out a single split into training, validation, and testing set which may cause some bias. We discuss this limitation in the end of the paper and propose the improvement in the future work.
Note that as the splitting ratio may influence the final results, we experiment four splitting ratio (training : validation : testing),  50\%:25\%:25\%, 60\%:20\%:20\%, 70\%:15\%:15\% and 80\%:10\%:10\% for each model respectively.

\begin{table}
\centering
\caption{Tag classification accuracy for four dataset splitting ratios in different methods.}
\label{tab:DatasetSplittingRatio}
\resizebox{1\textwidth}{!}{
\begin{tabular}{ccccccc} 

\toprule
 \textbf{Dataset splitting ratio} & \multicolumn{6}{c}{\textbf{ACCURACY} } \\ 
\cline{2-7}
\multicolumn{1}{l}{\begin{tabular}[c]{@{}c@{}}\textbf{(train:validation:test)}\\\end{tabular}} & \begin{tabular}[c]{@{}c@{}}\textbf{Histo }\\\textbf{ +SVM} \end{tabular} & \begin{tabular}[c]{@{}c@{}}\textbf{Histo }\\\textbf{ +DT} \end{tabular} & \begin{tabular}[c]{@{}c@{}}\textbf{ResNet }\\\textbf{ -pretraind} \end{tabular} & \begin{tabular}[c]{@{}c@{}}\textbf{ResNet }\\\textbf{ +pretraind} \end{tabular} & \textbf{Tag only}  & \textbf{Tag+Image}   \\ 
\hline
\textbf{50\%:25\%:25\%} & 0.6009 & 0.5531  & 0.6841 & 0.7410  & 0.7306 & \textbf{0.7940} \\
\textbf{60\%:20\%:20\%} & {0.6061} & {0.5629} & {0.6827} & {0.7402} & 0.7316 & {\textbf{0.7829}}\\
\textbf{70\%:15\%:15\%} & {0.6123} & {0.5805} & {0.6866} & {0.7387} & 0.7391 & {\textbf{0.7875}}\\
\textbf{80\%:10\%:10\%} & {0.5965} & {0.6249} & {0.7342} & {0.7545} & 0.7522 & {\textbf{0.8272}}\\

\bottomrule

\end{tabular}}
\end{table}

\subsection{Baselines}
We first set up several basic machine-learning baselines including the feature extraction (e.g., color histogram~\cite{wang2010robust}) with machine-learning classifiers (e.g., decision tree~\cite{quinlan1983learning}, SVM~\cite{cortes1995support}).
Color histogram (Histo) is a simple image feature that represents the distribution of colors in each RGB (red, green, blue) channel.
Based on these hand-crafted features, we adopt the support vector machine (SVM) and decision tree (DT) as the classifiers.
Apart from these conventional machine learning based baselines, we also set up several derivations of our model as baselines to test the importance of different inputs of our approach including with or without pretrained model (ResNet-pretained, ResNet+pretained), only using UI design image, existing tags or both of them (ResNet+pretained, Tag only, Tag+Image).
The training and testing configurations for these baselines are the same as that of our CNN model.

\begin{table*}
  \caption{Tag classification accuracy for five categories in different methods.}
  \label{tab:Result}
  \footnotesize
  \setlength{\tabcolsep}{0.3em}
  \begin{tabular}{lcccccc}
    \toprule
    \multirow{2}{*}{\textbf{CLASSES}} & \multicolumn{6}{c}{\textbf{ACCURACY}} \\ \cmidrule(r){2-7} & \textbf{\makecell{Histo+SVM}} & \textbf{\makecell{Histo+DT}} & \textbf{\makecell{ResNet-pretrain}} & \textbf{\makecell{ResNet+pretrain}} & \textbf{Tag only} & \textbf{Tag+Image} \\
    \midrule

    \textbf{App Function} &  &  &  &  &  & \\
    music & 0.6636 &{0.5545} & 0.6727 & 0.7909 & 0.8545 & \textbf{0.8909}\\
	food\&drink&{0.5765}&{0.6294} &0.7529 & {0.7882} & 0.7706& \textbf{0.8294}\\
	ecommerce & 0.5565 &{0.5726} &0.6895&  {0.7460} & 0.8306 & \textbf{0.8710}\\
	finance&{0.5655}&{0.5833}   &0.6964&   {0.7500} & 0.8274&\textbf{0.8512}\\
	travel&{0.5211}&{0.5842}    &0.7316&  {0.7053} & 0.8053 &\textbf{0.8474}\\
	game&{0.5814}&{0.5814}    &0.8062&   {0.7984} & 0.7597 &\textbf{0.8605}\\
	weather&{0.5745}&{0.7021}    &0.7447&   {0.7872} & 0.8085 &\textbf{0.8298}\\
	sport&{0.4220}&{0.6147}    &0.6147&  {0.6239} & 0.7064 & \textbf{0.7798}\\
	\cmidrule(r){0-0}
    
    \textbf{Color} &  &  &  &  &  & \\
	yellow&{0.5865}&\textbf{0.7596} & 0.7404 & {0.7404}&0.6442 &0.7500 \\
	red&{0.6667}&{0.7083} & 0.8194 & \textbf{0.8472}& 0.6111 &\textbf{0.8472}\\
	pink&{0.7609}&{0.6522} & 0.7826 & {0.7391} & 0.6522 &\textbf{0.8261}\\
	blue&{0.6600}&{0.6800} & 0.7700 & {0.7400}& 0.6800 & \textbf{0.8700}\\
	green&{0.7000}&\textbf{0.8714} & 0.8286 & {0.7714} & 0.6571 & {0.7857} \\
	white&{0.6111}&{0.6111} & 0.7778 & {0.7333}& 0.7333 & \textbf{0.7888} \\
	black&{0.6241}&{0.6015} & 0.8496 & {0.8271} & 0.6617 & \textbf{0.8571}\\
	\cmidrule(r){0-0}

	\textbf{Screen Function} &  &  &  &  &  & \\
	landing page&{0.5465}&{0.5346} &{0.7106}&{0.7017}& 0.7947 & \textbf{0.8115}\\
	signup&{0.4907}&0.5556&{0.7731}&{0.7130}& 0.7361 &\textbf{0.7778}\\
	checkout&{0.5545}&0.4182&{0.6000}&{0.7091}& 0.7545 &\textbf{0.8000}\\
	profile&{0.4667}&{0.5538}      &{0.5487}&{0.6513}&\textbf{0.9026} & 0.7590 \\
	\cmidrule(r){0-0}
	
	\textbf{Screen Layout} &  &  &  &  &  & \\
	dashboard & 0.5867 & 0.6067 &{0.7600}&{0.7933}& 0.7867 &\textbf{0.8800}\\
	chart&{0.6061}&{0.6667}     &{0.7121}&{0.7424}&\textbf{0.8485} & 0.8030\\
	form&{0.5429}&{0.5000}      &{0.6857}&{0.7429}& 0.5714 &\textbf{0.7714}\\
	list&{0.6136}&{0.5909}      &{0.7045}&\textbf{0.9091}& 0.6364 & 0.8182\\
	grid&{0.5000}&{0.5811}      &{0.6351}&{0.6486}& 0.7162 &\textbf{0.7432}\\
	\cmidrule(r){0-0}
	
	\textbf{Platform} &  &  &  &  &  & \\
	mobile&{0.7785}&{0.7847} & 0.8356 & {0.7954} & 0.9250 & \textbf{0.9402}\\
	website&{0.7513}&{0.7481} & 0.8418 & {0.8224} & 0.8837 & \textbf{0.9171}\\
	\midrule
	
	\textbf{Average} & {0.5965} & {0.6249} & {0.7342} & {0.7545} & 0.7522 & {\textbf{0.8272}}\\
    \bottomrule
  \end{tabular}
\end{table*}

\subsection{Evaluation Results}
\label{third:evaluation}
As we train a binary classifier for each tag, we adopt the accuracy as the evaluation metric for the model.
In different settings of data splitting, our model can still achieve the best performance compared with all baselines with average accuracy as 0.7875, 0.7829, 0.7940 (Table~\ref{tab:DatasetSplittingRatio}). 
It also demonstrates the robustness of our model in the scenario of the small training dataset.
Due to space limitations, we introduce more detailed performance only in one setting\footnote{More detailed results in other settings can be seen at \url{https://github.com/UITagPrediction/CSCW2020}} (80\% training, 10\% validation and 10\% testing) which can be seen in Table~\ref{tab:Result}. For comparison, the first 4 models (i.e., Histo+SVM, Histo+DT, ResNet-pretrain and ResNet+pretrain) only use images as input, while only textual information is used in ``tag only'' model. Our model (Tag+Image) combine both images and tags.
The traditional machine learning method based on the human-crafted features can only achieve about 0.6 average accuracy, while our model can achieve the average accuracy as 0.8272 i.e., the 32.4\% boost than the best baseline.
The good performance of our model is due to two factors.
First, the usage of the pretrained ResNet based on the ImageNet dataset increases the accuracy from 0.7342 to 0.7545.
Second, combining both the visual information and textual information can lead to better prediction accuracy than using any single information.

\begin{figure*}
	\centering
	\includegraphics[width=0.95\textwidth]{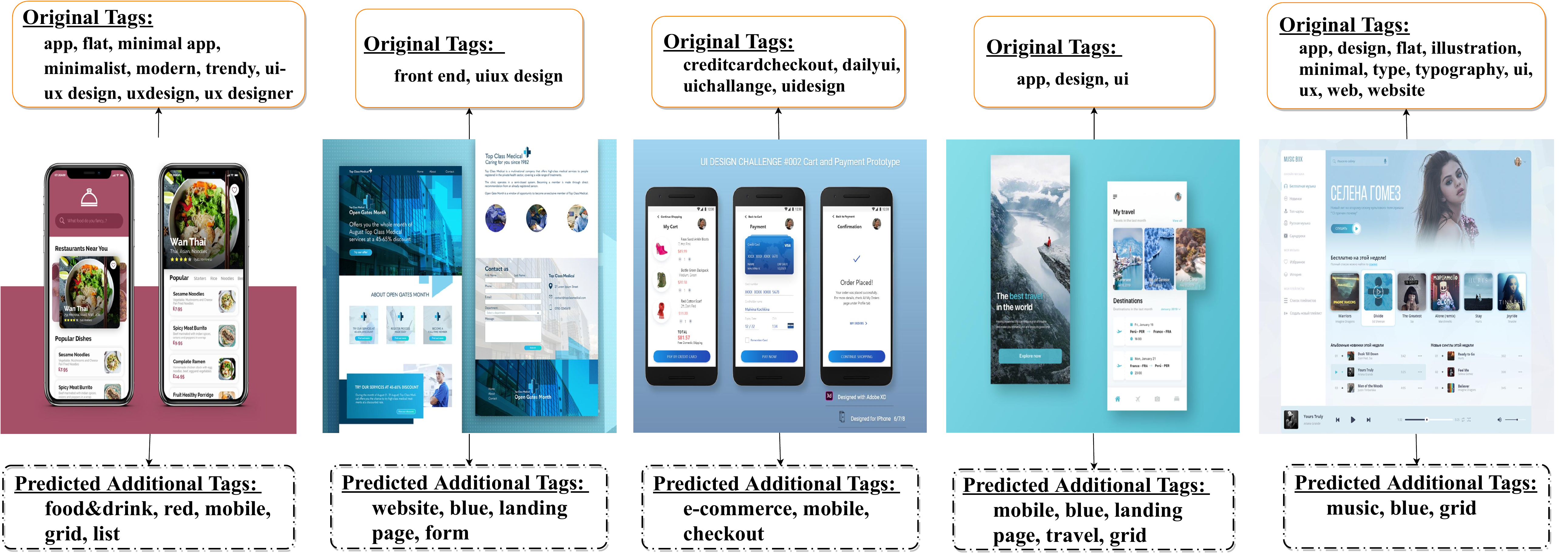}
	\vspace{-3mm}
	\caption{The predicted tags by our model for complementing the original tags.}
	\label{fig:UItagPrediction}
	\vspace{-0.8em}
\end{figure*}

\begin{figure*}
	\centering
	\subfigure[Predicted tag: game]{
		\includegraphics[width = 0.475\textwidth]{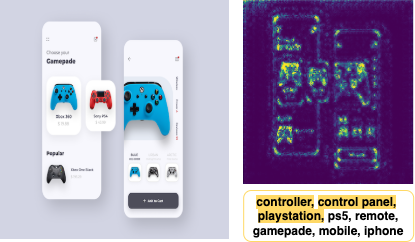}
		\label{fig:vis1}}
	\hfill
	\subfigure[Predicted tag: travel]{
		\includegraphics[width = 0.475\textwidth]{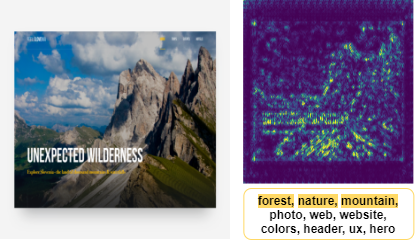}
		\label{fig:vis2}}	
	\hfill
	\subfigure[Predicted tag: music]{
		\includegraphics[width = 0.475\textwidth]{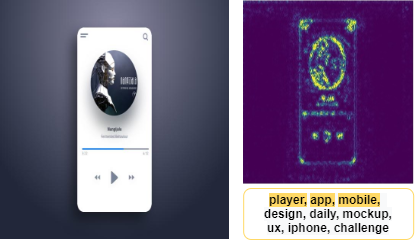}
		\label{fig:vis3}}	
	\hfill
	\subfigure[Predicted tag: chart]{
		\includegraphics[width = 0.475\textwidth]{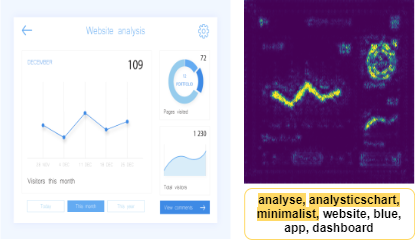}
		\label{fig:vis4}}
	\vspace{-3mm}
	\caption{Visualization of the salient features in our model leading to the final predictions.}
	\label{fig:deepvisualization}
	\vspace{-5mm}
\end{figure*}

Figure~\ref{fig:UItagPrediction} shows some predicted additional tags for example UI designs by our model.
It shows that our models can help locate the platform (e.g., ``website'' for the second UI), screen color (e.g, ``blue'' for the fifth UI), app functionality (e.g., ``food'' for the first UI), screen functionality (e.g., ``checkout'' for the third UI), and screen layout (e.g., ``grid'' for the fourth example).
All of these predicted tags are not appearing in the original tag set, and these additional tags can complement with the original ones for more effective UI retrieval.
Besides, once our method can be incorporated into the real site like Dribbble, our model can recommend users with potential tags before their design upload which proactively ensures the quality of design tags.
To show how our deep learning model works well for predicting the tag for the given UI design, we visualize the features learned in our model in Figure~\ref{fig:deepvisualization}.
For example, our model spots the play button which widely appears in the music player apps and highlights the "player" in the words in Figure~\ref{fig:deepvisualization}.(c) when predicting tag ``music''.
The visualization results reveal that our model does capture the visual and textual semantics of different UI designs. 



\begin{figure*}
	\centering
	\subfigure[No notable feature for the classifier to understand it correctly (missing "music" tag)]{
		\includegraphics[width = 0.3\textwidth]{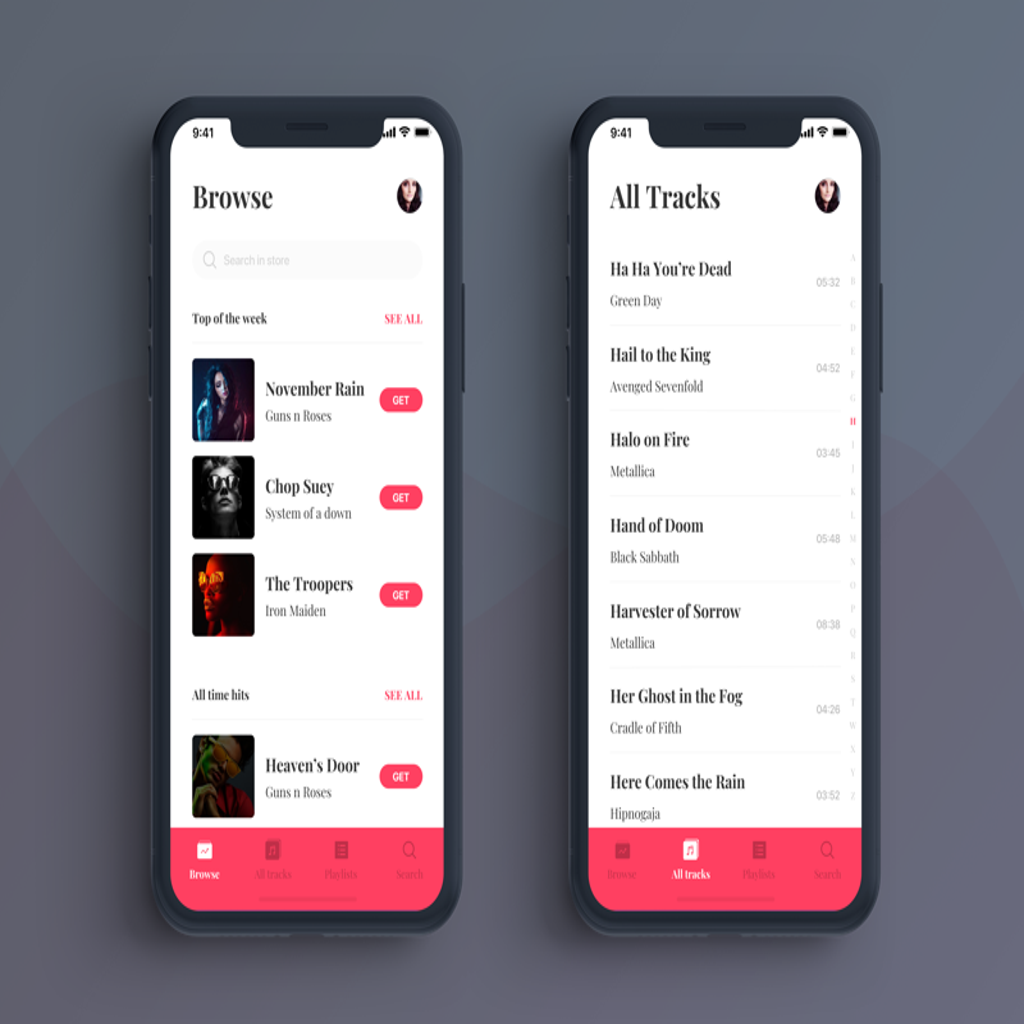}
		\label{fig:predicterror_1}}
	\hfill
	\subfigure[Some negative data is not really negative data (original design is tagged as "blue")]{
		\includegraphics[width = 0.3\textwidth]{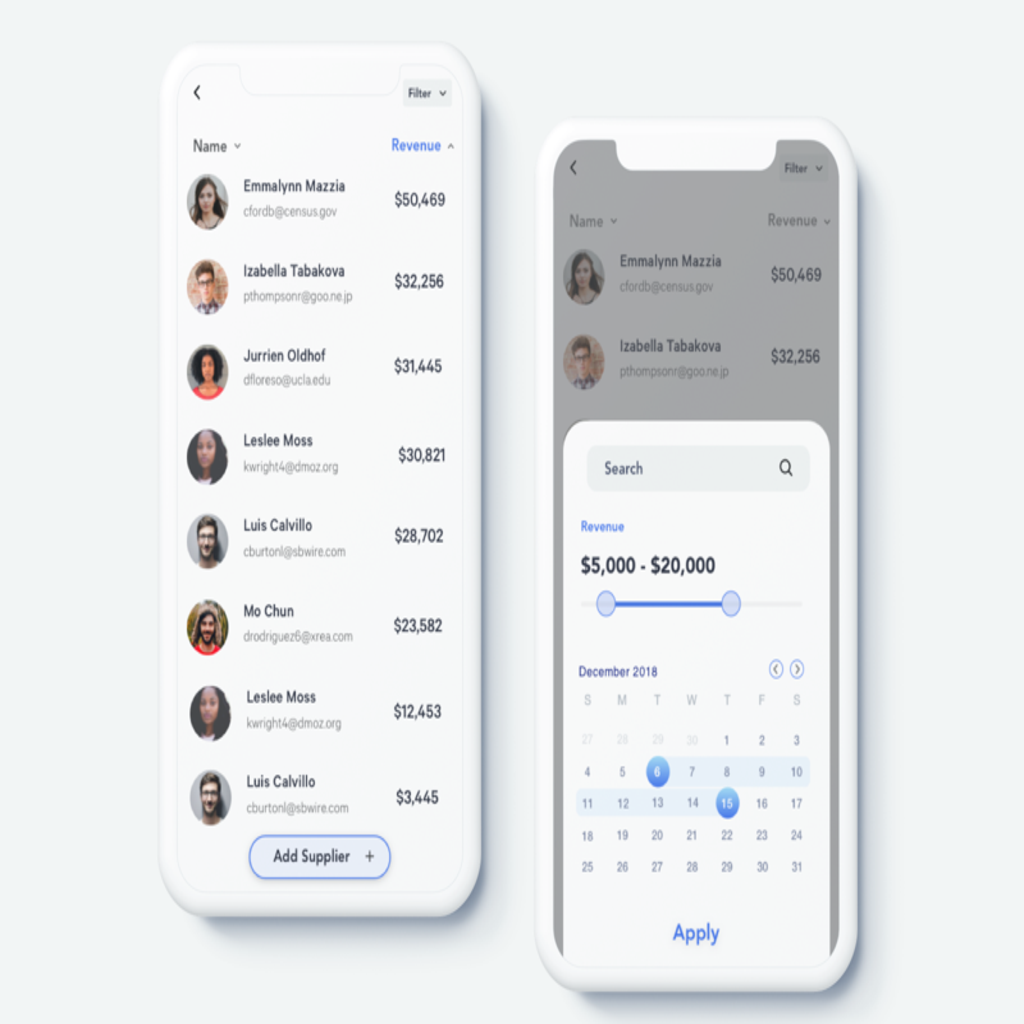}
		\label{fig:predicterror_2}}	
	\hfill
	\subfigure[Different preferences of designers (the design is tagged as "sport" while our model predict it as "e-commerce")]{
		\includegraphics[width = 0.3\textwidth]{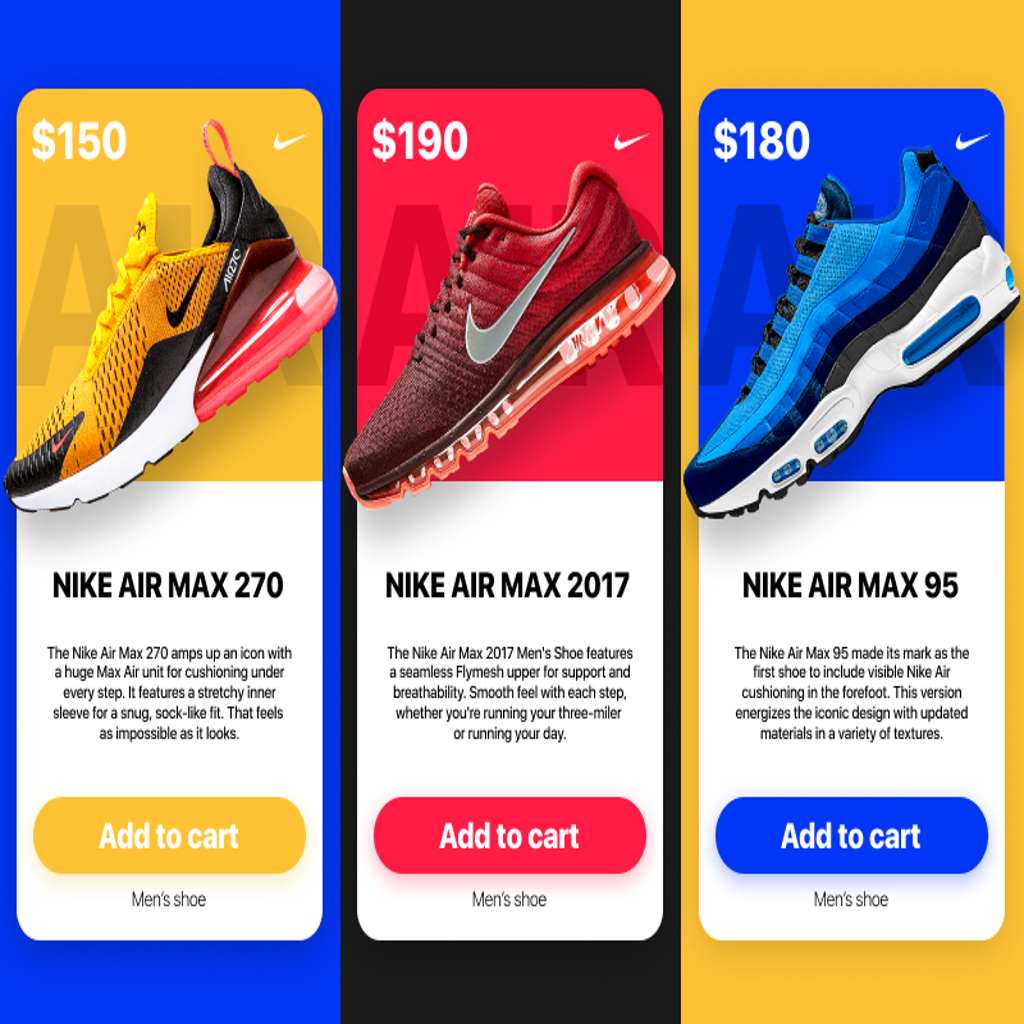}
		\label{fig:predicterror_3}}	
	\vspace{-5mm}
	\caption{Examples of the three kinds of prediction errors.}
	\label{fig:predicterror}
\end{figure*}

Albeit the good performance of our model, we still make wrong predictions for some UI designs.
To identify the common causes of prediction errors, we then further manually check the wrong predictions in the test dataset.
According to our observation, there are two main reasons leading to the wrong prediction.
First, some UI designs are too complicated for the model to recognize the right tag for it. 
For example, the design in Figure ~\ref{fig:predicterror_1} is tagged with a ``music'' class. 
However, there is no notable feature for our classifier to understand it correctly.

Second, some negative data is not really negative data due to the different preferences of designers.
Our way to collect negative data assumes that one UI design image that does not contain the target tags but contains other same-category tags should not be attached to the target tag.
Although it works for most cases, the collected negative data still contains some ``noisy'' UI designs i.e., UI design not attached with the target tags (maybe due to the designers' neglect) should be attached with tag.
As seen in Figure~\ref{fig:predicterror_2}, although the main color is white, the designers still assign the tag ``blue'' to highlight the selected date in the calendar.
It also applies to Figure ~\ref{fig:predicterror_3} whose label is ``sport'' while our model recognizes it as the ``e-commerce'' app.
Although our prediction does not match the label, it still makes sense.

\section{Retrieval Evaluation}

The experiments in the last section demonstrate the accuracy of our model.
We conduct a pilot user study in this section to further evaluate the usefulness of the predicted addition tags for bootstrapping the UI design retrieval. 


\subsection{Procedures for User Study}
The number of words in each query is 2.35 on average, according to an analysis of an AltaVista search engine query log for almost 1 billion search requests from approximately 285 million users over six weeks~\cite{silverstein1999analysis}. 
Therefore, based on the five main categories of tags in Table~\ref{tab:tagCategory}, we randomly selected three tags from three categories respectively as the query.
To mimic the real queries from designers, the query contains one platform tag and two tags from the other two random categories.
We obtain 5 queries as seen in Table~\ref{tab:query}.

We set up two groups for each query, i.e., control group and experimental group. 
The query database contains 61,700 UI designs collected from Section~\ref{sec:empiricalStudy}.
For the control group, we directly retrieved the UI without tag augmentation in the database, which is regarded as a baseline for comparison.
For experimental group, we searched the UI design database in which UI tags have been normalized by using the extracted abbreviations and synonyms in Section~\ref{sec:tagNormalization} and complemented with additional predicted tags as described in Section~\ref{sec:tagAugmentation}. 
Note that the UI can only be retrieved if it contains all keywords in the query.
As there may be too many results from the experimental group (as seen in Table~\ref{tab:query}), we randomly take 10, 30, 50 retrieval candidates as the experimental groups so that it is fair to compare with the control group.

We recruit 10 Master and final-year Bachelor students from our school, and each of them is assigned the results of five queries from one control group and three experimental groups. 
Note that they do not know which result is from the experimental or control group.
Given each query, they individually mark each result as related to the query or not and for each query, we will randomly shuffle the order of candidates to avoid potential bias.
After marking results for each query, they are asked to rate how satisfied they are with the retrieval results in a five-point likert scale (1: not satisfied at all and 5:highly satisfied), and how diverse also in five-point likert scale.

\begin{table*}
  \caption{The random queries for searching UI designs}
  \vspace{-3mm}
  \label{tab:query}
  \begin{tabular}{llcc}
    \toprule
    \bf{ID} & \bf{Query}  &\textbf{\#UI (control group)} & \textbf{\#UI (experimental)}\\
    \midrule
    1 &  iphone+e-commerce+checkout    & 3 & 604 \\
	2 &  mobile+food\&drink+listview   & 4 & 510\\
	3 &  website+blue+landing page     & 9 & 1838\\
	4 &  ios+finance+chart             & 4 & 1018\\
	5 &  web+sign up+form              & 10 & 440\\ 
    \bottomrule
  \end{tabular}
\end{table*}

\begin{table*}
  \caption{The comparison of the experiment and control groups. $^*$denotes $p<$ 0.01, $^{**}$denotes $p<$ 0.05.}
  \vspace{-3mm}
  \label{tab:score_Tabel}
  \begin{tabular}{lcccc}
    \toprule
    \textbf{Measures} & \textbf{Control} & \textbf{Experiment 10} & \textbf{Experiment 30} & \textbf{Experiment 50}\\
    \midrule
    \#Related Candidate & 3.2/6 (53.3\%) & 7.5/10 (75.6\%)$^{**}$ & 25.2/30 (84.1\%)$^*$ & 45/50 (90\%)$^*$\\ 
	Satisfaction score & 2.8 & 3.56$^{**}$ & 4.28$^{**}$ & 4.8$^*$\\ 
	Diversity score      & 2.22 & 2.68$^*$ & 4.16$^*$ & 4.7$^*$\\ 
    \bottomrule
  \end{tabular}
  \vspace{-3mm}
\end{table*}

\subsection{Results of Retrieval Effectiveness}
Table~\ref{tab:score_Tabel} shows that the experimental group retrieves many more related candidates than the control group in both the absolute the number and relative percentage (75.6\%, 84.1\%, 90\% compared with 53.3\%). 
That is the biggest strength of our approach i.e., recovering the missing tags, resulting in more relevant UI designs given the query.
Compared with the baseline, most participants admit that our method can provide more satisfactory and diverse results for them to choose from.
And with the increase of the candidate number in the experimental group, participants give higher satisfaction and diversity scores (3.56, 4.28, 4.8 in satisfaction score, and 2.68, 4.16, 4.7 in diversity score) which are significantly higher than that (2.8, 2.22) of the control group. 
It is consistent with human perception, as more results lead to a higher possibility to see the serendipitous idea. 
The detailed results can be seen in our site\footnote{\url{https://sites.google.com/view/uitagpredictionuserstudy/home}}.

To understand the significance of such differences, we carry out the Mann-Whitney U test~\cite{fay2010wilcoxon} (specifically designed for small samples) on the number of useful candidate, satisfactory, and diversity ratings between the experiment and the control group respectively. 
The test results suggest that our method does significantly outperform the baseline in terms of these metrics with $p < 0.01$ or $p < 0.05$.
Although by no means conclusive, this user study provides initial evidence of the usefulness of our method for enhancing the performance of the tagging-based search.

\section{Discussion}
\label{sec:discussion}
In this section, we discuss the implication of our work on designers, design sharing platform, and the generalization of our approach.

\textbf{On designers:} Designing good UI is a challenging task and the design taste of users is changing dynamically. 
To continually follow the trend, designers always resort to the online sharing resources such as design kits, blogs, design sharing websites for inspirations.
Apart from browsing others' designs, many designers also actively share their own designs into the platform.
Despite that the generosity is appreciated, more meta information related to the UI is also welcomed.
The precise and complete tags attached to the UI can attract more designers, potentially leading to the popularity and reputation of you and your design. 
So, it is necessary for designers to add as many and precise tags to their design before uploading.

\textbf{On design sharing platform:} To ensure the quality of UI design tags for supporting effective retrieval, the design sharing platform should develop a proactive mechanism. 
Although our model can recommend missing tags for indexing the UI semantics which further benefits the community with textural retrieval and management, there are still some limitations using it solely.
First, without the support of the design platform, we can only implement our model as an extension of the browser like Chrome.
Then it works as a reactive way for adding tags to the existing UI design, but missing tags may already harm the designers who want to search before our tag recommendation.
Second, although our model can achieve good accuracy, there are still some error predictions that need some human efforts to confirm the recommended tags. 

Therefore, in addition to the reactive addition of tags to the existing UI design from our side, we also need a more proactive mechanism of tag assurance which could check a UI design before it is uploaded by spotting the potential missing tags, and reminding the design owner to confirm the recommended tags and add them. 
The simplest way is to list the vocabulary of existing tags aside when designers are filling in tags for their sharing design.
We also hope that our model can be embedded into the design sharing sites, behaving as a proactive mechanism to not only automatically recommend designers with potential tags for choosing, but also warn them about wrong tags they adopt initially such as inconsistent tagging vocabulary (i.e., morphological forms).

We can also leverage the crowd source i.e., the peer designers in the community to ensure the completeness and correctness of tags, and that mechanism is widely used in content generation sites like Wikipedia and Stack Overflow.
The design sharing site should also allow and encourage moderators to edit community-generated tags.
Previous studies~\cite{li2015good, chen2017community, choi2018will} indicate that experienced moderators are able to improve content consistency and quality by making lasting edits to low-quality posts, especially content curated by novices in Q\&A sites.
There may be extra benefits to emphasize the role moderators can play at standardizing content across the site.

\textbf{On the generalization of our approach}: We report and analyze the performance of our CNN-based model for predicting missing UI tags in Section \ref{third:evaluation}. 
One limitation with our model is that we current only test our model on 26 tags with enough corresponding UI designs. 
But note that given a new tag, our model just needs to collect all GUI designs with/without that tags as the training data after normalizing the tags. We then can train a specific binary classifier for amending that tags to the GUI design.
Another limitation with our evaluation is that due to the time limit, we only take a single split of the dataset to plot the evaluation accuracy which may bring some noise or bias.
In the future, we will try to mitigate that effect by adopting K-fold cross evaluation.
In addition, our approach is not limited to UI designs. 
In fact, we do similar experiments in other tag categories with high frequency and large instances (20000 $\sim$ 60000 images) in Dribbble, including tag ``illustration'', ``icon'' and ``logo''. 
Table \ref{tab:ResultsForNonUI} shows that our model also achieves high accuracy in non-UI-related tags. The result demonstrates that our deep-learning method can be generalized to a broader range of graphic designs.


\begin{table*}
  \caption{Tag classification model accuracy for non-UI-related tags}
  \vspace{-3mm}
  \label{tab:ResultsForNonUI}
  \begin{tabular}{lc}
    \toprule
    \textbf{Classes} & \textbf{Tag+Image accuracy}\\
    \midrule
    illustration  & 0.9336 \\
	icon   & 0.9296\\
	logo   & 0.9478\\
    \bottomrule
  \end{tabular}
  \vspace{-5mm}
\end{table*}

\section{Conclusion \& Future Works}
UI is crucial for the success of the software especially for mobile apps in the market.
Albeit a large number of UI designs online with human tags, many UIs are still missing some closely related keywords for retrieval.
To assist designers with better UI design retrieval, we adopt a deep learning method to automatically recommend additional tags that convey the semantics of the UI.
Based on the systematical analysis of the existing large-scale UI designs, our model can achieve the accuracy as 0.827 which significantly outperforms other baselines.
The user study also demonstrates the usefulness of our model in UI retrieval.

In the future, we are extending our research works into three different directions.
On the one hand, we are diving into the model to improve the current accuracy performance.
Currently, we only evaluate our model on tags which have many related UI as the training data, and we will also customize our model for achieving good performance on tags with only small datasets.

On the other hand, we are exploring the searching for dynamic animation UI designs.
According to our observation, apart from the static UI design images, there are also some animation designs in GIF format which show how users interact with the UI dynamically.
We will leverage the video analysis method to analyze the human actions and design interactions within those GIFs, so that designers can search the animation with textual queries. 

Furthermore, to understand the usefulness of our technology in a real design context, we hope to cooperate with professional designers and design sharing sites by deploying our model. We will explore how designers actually make use of the retrieved UIs, and collect valuable feedback to improve our technology and evaluation methods.





\begin{acks}
	We appreciate the initial exploration by Ruiqi Wang of this work.
	This project is partially supported by Facebook gift funding and ANU-Data61 Collaborative Research Project CO19314.
\end{acks}

\bibliographystyle{ACM-Reference-Format}
\bibliography{reference}


\end{document}